\documentclass{nature}
\usepackage{graphicx} 
\usepackage{xcolor}

\title{
Thermal transport of helium-3 in a strongly confining channel}

\author{D. Lotnyk$^1$, A. Eyal$^{1,2}$, N.~Zhelev$^{1}$, T.S.~Abhilash$^{1,3}$, E.N.~Smith$^1$, M.~Terilli$^1$, J.~Wilson$^{1,4}$, E. Mueller$^1$, D.~Einzel$^5$, J.~Saunders$^6$ and J.M.~Parpia$^1$
}
\makeatletter 
\let\saved@includegraphics\includegraphics 
\AtBeginDocument{\let\includegraphics\saved@includegraphics} 
\renewenvironment*{figure}{\@float{figure}}{\end@float} 
\makeatother 

\begin{document}
\maketitle

\begin{affiliations}
 \item Department of Physics, Cornell University, Ithaca, NY, 14853, USA
 \item Physics Department, Technion, Haifa, Israel
 \item VTT Technical Research Centre of Finland Ltd, Espoo, Finland
 \item SUNY Geneseo, Geneseo, NY, 14454, USA
 \item Walther Meissner Institut, Garching, Germany
 \item Department of Physics, Royal Holloway University of London, Egham, TW20 0EX, Surrey, UK
\end{affiliations}
\today{}

\begin{abstract}
\begin{center}
    Abstract:
\end{center}

In a neutral system such as liquid helium-3, transport of mass, heat, and spin provide information analogous to electrical counterparts in metals, superconductors and topological materials.  Of particular interest is transport in strongly confining channels of height approaching the superfluid coherence length, where new quantum states are found and excitations bound to surfaces and edges should be present.  Here we report on the thermal conduction of helium-3 in a 1.1~$\mu$m high microfabricated channel. In the normal state we observe a diffusive thermal conductivity that is approximately temperature independent, consistent with recent work on the interference of bulk and boundary scattering. In the superfluid state we measure diffusive thermal transport in the absence of thermal counterflow. An anomalous thermal response is also detected in the superfluid which we suggest may arise from a flux of surface excitations.  

*Correspondence to: jmp9@cornell.edu
\end{abstract}

\section*{Introduction:}

Confined superfluid $^3$He is an ideal model system for studying topological quantum matter\cite{Levitin13PRL,Levitin13Science}. The predicted surface and edge excitations are expected to be measurable in thermal transport\cite{Volovik2009,QuiZhangRMP2011,Sauls11PRB,Nomura12} in $^3$He. Moreover, thermal analogues of the Hall effect are predicted in the $^3$He chiral A-phase\cite{Volovik1988} both as a result of edge currents and scattering from impurities. Observation of these exotic phenomena requires the confinement of $^3$He in precise geometries. However, as yet there are no measurements of the thermal conductivity of superfluid $^3$He under confinement even in nominally more conventional regimes. 
Here we report that thermal transport in confined channels is rich with unanticipated effects in both the normal and superfluid states. Quantifying this transport provides insight into the underlying kinetic processes and paves the way for distinguishing the signatures of topological superfluidity.

Experiments on $^3$He in the presence of disorder have shown that in addition to the modification of transport behavior from the pure liquid\cite{Dmitriev2015,SaulsSharma10,EinzelPRL98,Fisher03}, new superfluid phases emerge due to the anisotropy of the disorder\cite{DmitrievPRL15,SaulsPRB13,HalperinNatPhys12,HalperinNatPhys13,ZhelevNC2016,Dmitriev2010}. These anisotropic structures have also led to the observation of half-quantum vortices\cite{Mineev2014,AuttiPRL17}. Nanofabrication techniques can be used to engineer anisotropic environments\cite{SaulsWimanPRB15,Wiman2014}, with no accompanying disorder. Simple confinement in a slab can stabilise new superfluid phases, such as the recently observed spatially modulated superfluid\cite{VorontsovPRL07,LevitinPRL19}. More complex structures, such as channels or periodic arrays of posts, with typical length scales of a few coherence lengths, can also potentially tailor new superfluid phases\cite{Wiman2014}. Thermal transport will play a key role in characterizing these new ``materials” and hybrid structures. In the work presented here, the focus is on the understanding of thermal transport of $^3$He in a simple slab geometry with strong confinement corresponding to of order 15 to 50 times the pressure dependent zero temperature coherence length, $\xi_0=\hbar v_F/(2\pi k_B T_c)$ where $v_F$ is the Fermi velocity, $k_B$ Boltzmann's constant, and $T_c$ the superfluid transition temperature.

Thermal conductivity in normal $^3$He is a diffusive process and can be understood in terms of the kinetic theory of quasiparticle excitations which collide, exchanging and transmitting energy.  Due to the Pauli exclusion principle, the phase space available for scattering becomes small at low temperatures giving rise to a strong temperature dependence of the inelastic thermal mean free path, $\lambda_{\kappa} \sim$ $T^{-2}$. This results in a bulk thermal conductivity, $\kappa$ that increases as $T^{-1}$ (since $\kappa = 1/3 (C_v/V)v_F\lambda_{\kappa}$, where $V$ is the molar volume, $C_v \sim T$ is the molar specific heat, $v_F$ is the Fermi velocity\cite{Greywall84TC}).  This behaviour is observed in the bulk liquid, since, unlike other condensed matter systems, $^3$He is impurity-free and there are no elastic scattering centers.  Introduction of a collection of point scatterers (such as aerogel) leads to a vanishing conductivity\cite{Reeves02,Sauls00,Choi07} as T $\rightarrow$ 0, due to the mean free path being limited by scattering from the impurities. Recently, there has been significant renewed interest in hydrodynamic transport in electron fluids, arising from advances in materials. Building on early work on two dimensional electron gases in AlGaAs heterostructures\cite{MolenkampPRB1995}, the required condition that electron-electron collisions dominate over electron-phonon or electron-impurity scattering is satisfied in ultraclean materials such as graphene\cite{BandurinNC2018}, PdCoO$_2$\cite{MollScience2016}, and WP$_2$\cite{GoothNC2018} leading to viscous and quasiballistic transport with signatures distinct from Ohmic transport. The confinement of such materials into restricted conduction channels is relevant for the understanding of the interplay of bulk and surface scattering, with many open questions. In this context, confined $^3$He, in which scattering between quasiparticles dominates in bulk, provides a useful paradigm, including the potential to cross-over to quasi-two-dimensional transport\cite{Sharma11PRL} with strong confinement.

In our experiment two chambers filled with bulk fluid, a small isolated volume and a container with a heat exchanger through which the $^3$He is cooled, are
separated by a nanofabricated 1.1 $\mu$m high channel. Both containers are equipped with a tuning-fork thermometer which can measure the temperature or act as a heater. By injecting heat into one chamber and measuring the response we explore the $^3$He diffusive thermal conductivity under strong confinement in both the normal and superfluid phase.

In the normal state we find an anomalous thermal conductivity that is nearly temperature independent below 10 mK, implying an effective mean free path that varies as $T^{-1}$. This is the same temperature dependence as that of the momentum relaxation time inferred from our earlier mass transport studies in $^3$He films on polished silver surfaces\cite{Casey04,Sharma11PRL}. We suggest that these results may be accounted for by quasi-classical interference between bulk scattering and that arising from surface disorder\cite{MeyerovichPRB1998,MeyerovichPRB2002}.

In addition to diffusive heat flow\cite{Einzel84JLP}, superfluids support thermal transport via a hydrodynamic process: two-fluid counterflow where relative motion of the superfluid and normal component results in heat flow. This effect is well established\cite{Allen38,Daunt39,Kapitza41,London38,London39} in studies of superfluid $^4$He, but results on superfluid $^3$He are limited\cite{Johnson75,Kleinberg74}. In steady-state thermal counterflow through a channel, the temperature gradient generates a fountain pressure, such that the difference in chemical potential between the two ends of the channel is zero. The superfluid component (driven by gradients in chemical potential) flows at constant velocity towards the hot end. The fountain pressure forces the normal (entropy carrying) component in the opposite direction, with volume flow rate determined by viscous transport in the channel. Near $T_c$ this hydrodynamic thermal transport dominates if the normal component is not viscously clamped.

One motivation of the present experiment was to quantify the diffusive thermal transport in the superfluid phase, arising from quasiparticle excitations, by reducing thermal counterflow. Informed by prior mass-flow studies\cite{EinzelParpia87,Casey04,Sharma11PRL}, the strong confinement imposed by the 1.1 $\mu$m channel was designed to clamp the normal component even in the presence of slip of the normal component in the extreme Knudsen regime. The Knudsen regime onsets when the viscous mean free path exceeds the height of a confining channel and is accompanied by slip (the phenomenon where the velocity of a fluid in contact with a wall is non-zero) that allows the viscous fluid to move relative to the wall. We find that the conditions are met so that the contribution from hydrodynamic flow is negligible. The measured diffusive thermal conductivity under confinement shows a weak temperature dependence similar to theoretical predictions for thermal conductivity in bulk superfluid.

However, in the superfluid state we observe a further unexpected response of the thermometer in the heat exchanger volume. The non-local response is indicative of a non-equilibrium effect in the thermal transport. Although the overall length of the channel is several mm, much longer than the inelastic mean free path of bulk quasiparticle excitations, the observed response appears to indicate a ballistic flow of quasiparticles induced by the fountain pressure created in the heated isolated volume, for which surface bound excitations may be responsible\cite{ZhangPRB1987,ThunebergPhysica1992,Nagato1998,VorontsovPRB2003}.

\begin{figure}[t]
\includegraphics[width=\textwidth]{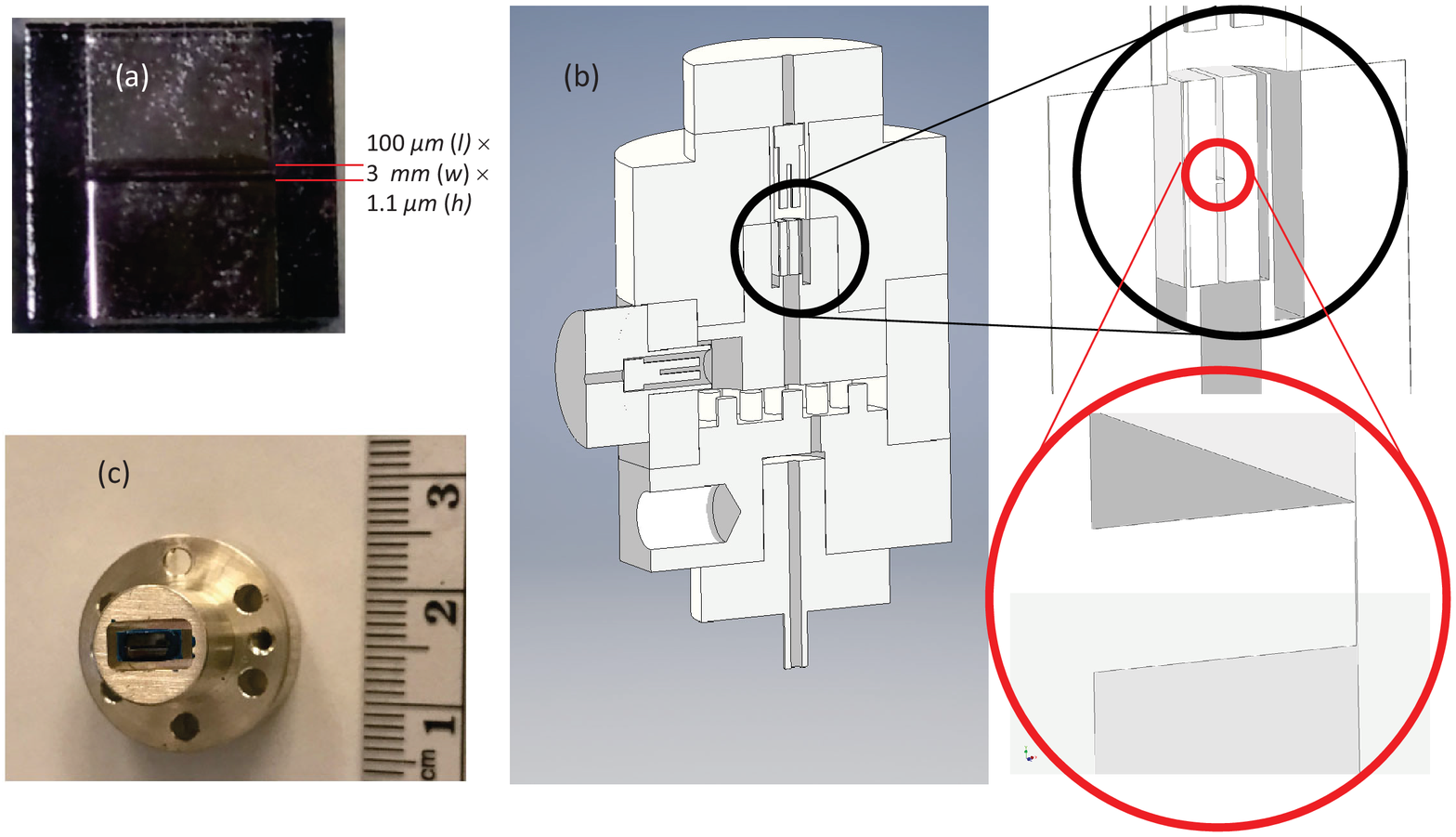} 
\caption{
\label{fig::1_Details of Experimental Cell}
  (a) Image of the cavity containing the channel prior to mounting. (b) Schematic of the experimental cell with forks mounted in the isolated chamber (IC depicted at the top) and in the chamber containing the heat exchanger (HEC located in the chamber below the channel) separated by the thermal conductance channel. The cavity in its mount is depicted schematically in the large black circle and the channel is depicted in the large red circle. (c) Cavity mounted in coin-silver carrier, where the (blue) epoxy joint to the thin coin-silver wall is visible. 
  }
\end{figure}

\section*{Experimental details.}
This paper describes the measurement of heat transport through a channel which is 1.1~$\mu$m high, 3~mm wide, and 100~$\mu$m long with 200~$\mu$m tall $\times$ 3~mm wide $\times$ 2.45~mm long ``lead-in" sections defined at either end of the main channel. The design is such that the 1.1 $\mu$m height section should dominate the thermal impedance. The structure is shown in Figure~\ref{fig::1_Details of Experimental Cell}(a,b). Two chambers sit on either end of this channel, one of which is thermally anchored to the nuclear demagnetization stage\cite{Parpia84RSI} through a sintered silver heat exchanger. We refer to this as the Heat Exchanger Chamber (HEC) (Figure~\ref{fig::1_Details of Experimental Cell}(b)).  The second chamber (designated as the Isolated Chamber, IC) was cooled through the thermal impedance provided by the channel, or, at temperatures above $\sim$10~mK by direct thermal contact with the coin-silver walls via the Kapitza thermal resistance\cite{Cousins94}. The channel was nanofabricated in 1~mm thick silicon, capped with 1 mm thick sodium doped glass, anodically bonded to the silicon\cite{Zhelev18RSI,Wallis69}. The channel was glued into a coin silver carrier (Figure~\ref{fig::1_Details of Experimental Cell}(c)) using epoxy\cite{Trabond}. The temperature in each chamber was determined by a quartz ``tuning-fork" thermometer operating at 34 kHz.\cite{Blaauwgeers07}.

A heat pulse was applied to the liquid in the IC by increasing the drive voltage applied to the fork in the IC by up to a factor of ten for a period of 10 to 100 seconds in the superfluid and 60 to 300 seconds in the normal state. These pulses deposited energy of order a few nJ compared to the ambient power dissipated by the fork of order 0.1~pW. The drive was then restored to the usual level and the quality factor of the fork, $Q$ (and hence the temperature of the IC) was monitored through its recovery to determine the thermal relaxation time. The measured IC fork thermal relaxation time, $\tau$  was then related to the thermal resistance, $R_{th}$ through $\tau = R_{th} C$ where $C$ is the heat capacity of the $^3$He in the IC volume. The heat capacity was determined from the known specific heat of $^3$He\cite{Greywall86SH}, and the calculated volume of the isolated chamber (0.14$\pm$0.02~cm$^{3}$), shown in ~Figure \ref{fig::1_Details of Experimental Cell}. The geometry was chosen so that the anticipated thermal relaxation times\cite{Greywall84TC,Greywall86SH} would lie between 100 and 3,000 seconds, compatible with the response time of the tuning fork thermometer. The equilibrium temperature of the $^3$He sample was also determined by monitoring the $Q$ of the HEC tuning fork.

Data was obtained while warming and cooling the nuclear demagnetization stage\cite{Parpia84RSI} to which the cell was thermally anchored. To cancel out the ambient heat leak to the nuclear stage of a few nW, we swept the magnetic field at a rate close to that needed to maintain a constant temperature. Thus we could achieve a linear temperature ramp while warming or cooling. The temperature was monitored with a melting curve thermometer (designated $T_{MCT}$)\cite{Greywall82JLTP,Greywall85MCT} anchored to the nuclear stage. In practice the temperature ramp of 15-35~$\mu$Khr$^{-1}$ was slow enough to allow temperature sweeps from 0.3 $T_c$ (the lowest temperature achievable in the liquid with our nuclear stage) to above $T_c$ in about 2 days (or the reverse). This time was sufficient to apply $\sim$ 50 pulses in a warm-up or cool-down permitting ample time for thermal recovery between pulses. Measurements close to $T_c$ were carried out with slower temperature ramps, since the thermal time constants were longer in the vicinity of $T_c$.
Measurements in the normal state were carried out up to 100~mK. The magnetic field on the nuclear stage was used to vary the temperature (imposing a linear magnetic field ramp) usually while warming, even up to 100 mK.

\begin{figure}[t]
\includegraphics[width=\textwidth]{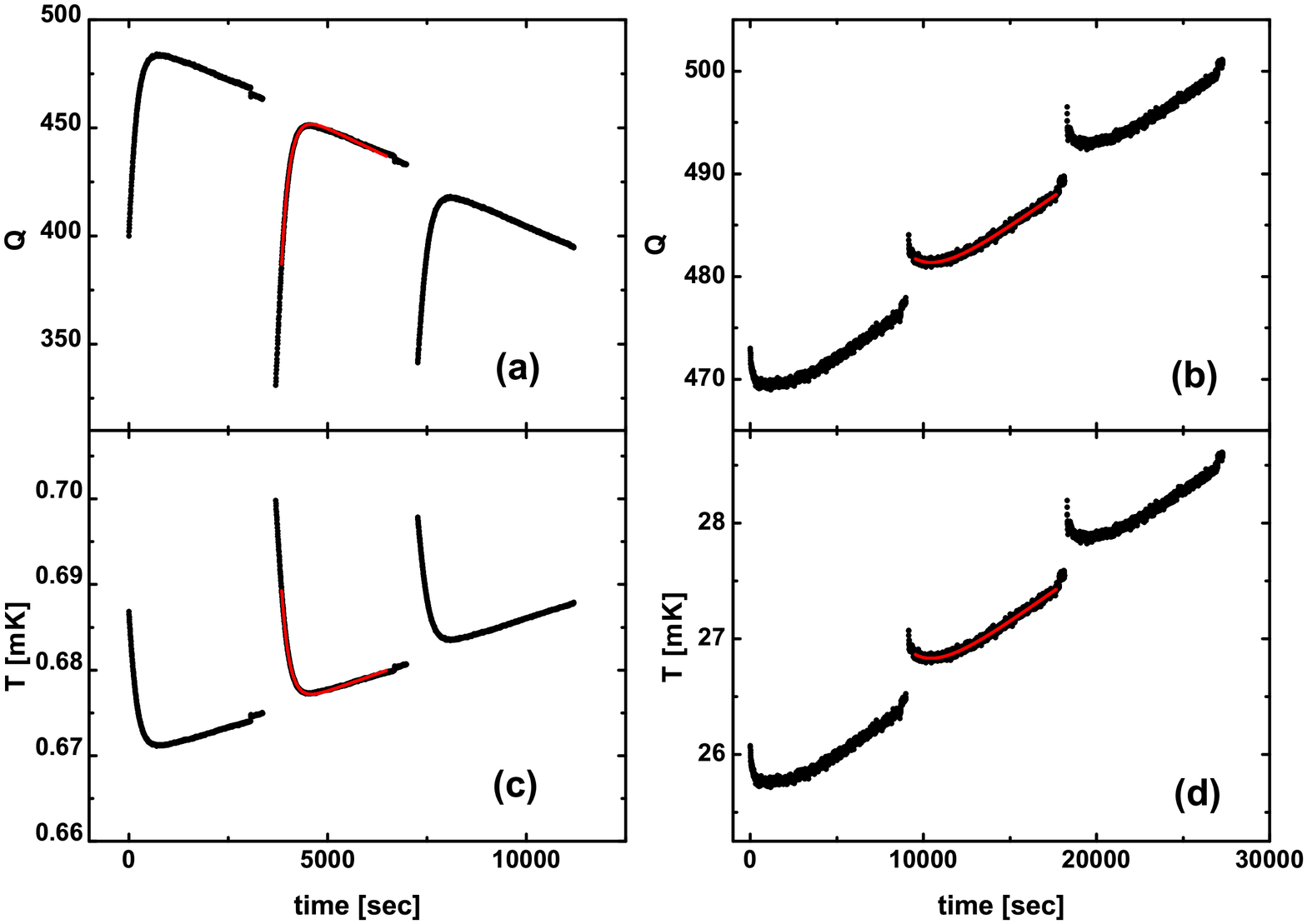} 
\caption{\label{fig::2_Heat Pulses}
  (a) Typical pulses applied in the superfluid state recording the $Q$ vs time and (b) in the normal state at 22 bar. Also shown in red are the fitted decays to an exponential with additional linear term to account for the steady temperature drift. The $Q$ response is reversed in the normal and superfluid states, because the viscosity decreases with increased temperature above $T_c$ while it decreases below $T_c$. (c, d) show the corresponding inferred temperature responses.}
\end{figure}

We measured the resonant frequency, $f$, and the quality factor, $Q$, of both quartz tuning forks as a function of temperature. The $Q$ of the HEC fork was calibrated against $T_{MCT}$ during slow temperature sweeps at each pressure in the absence of pulses. Since the IC and HEC forks were nearly identical, they display similar characteristics and the $Q$ vs. $T$ calibration was transferred from the HEC to the IC fork after correction for the difference in $Q^{-1}$ at $T_c$. Both quartz forks were operated in digital feedback loops (see Methods to follow) and maintained near their resonant frequencies. Typical responses to applied heat pulses to the IC fork are shown in Figure~\ref{fig::2_Heat Pulses}(a,b) both below and above $T_c$ at 22~bar. The relaxation responses above and below the superfluid transition are inverted due to the opposite temperature dependence of the viscosity in the normal and superfluid states. The transient following a heat pulse was fit to an exponential recovery along with a linear term to account for the temperature drift. Temperature excursions from ambient were limited to a few percent and are illustrated in Figure~\ref{fig::2_Heat Pulses}(c,d).

\begin{figure}[t]
\begin{center}
\includegraphics[width=0.6\textwidth]{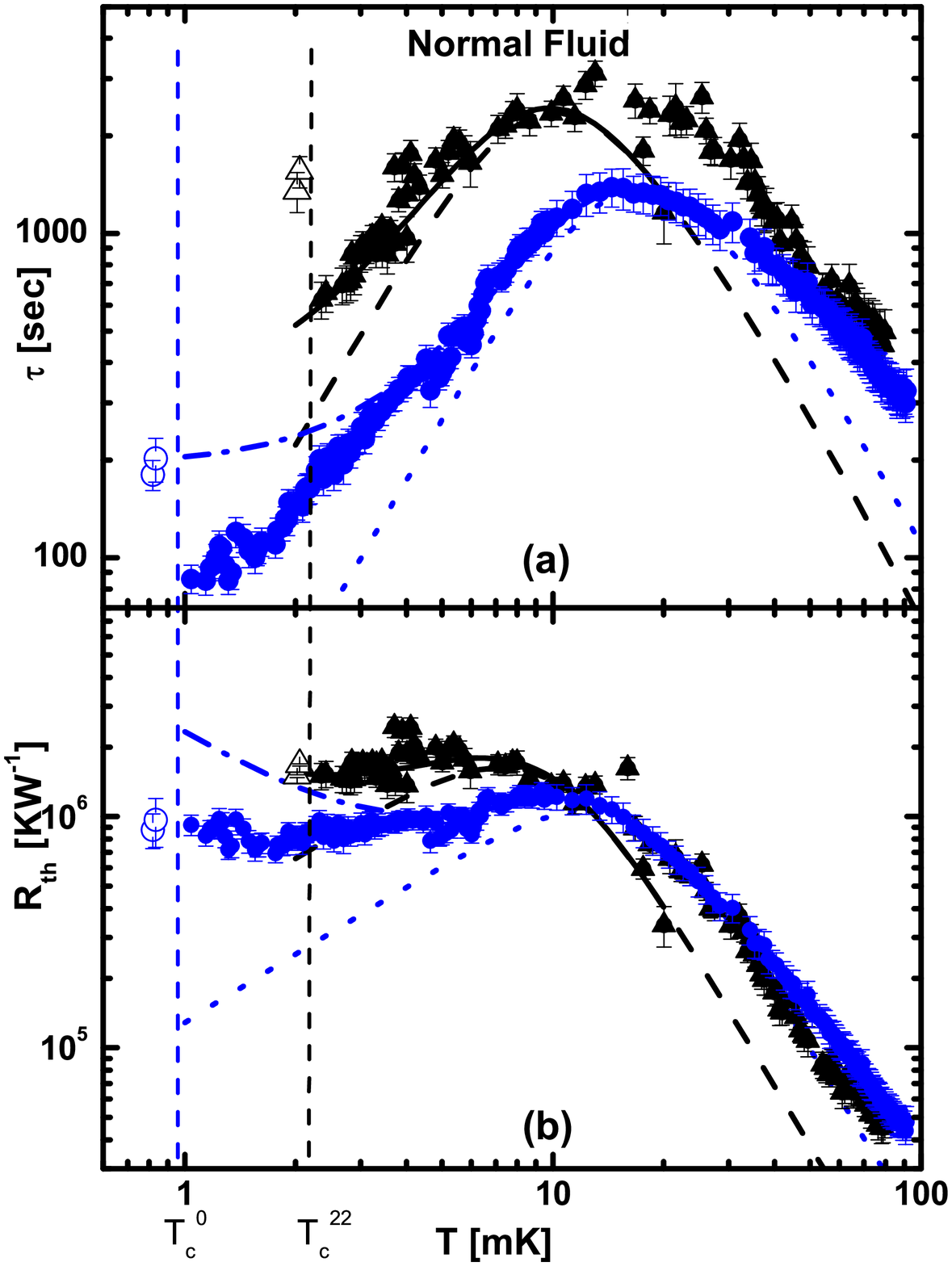} 
\end{center}
\caption{\label{fig::3_NormalState}
\small
  The measured thermal relaxation times (a) and calculated thermal resistances (b) at 0~bar (blue circles) and 22~bar (black triangles) in the normal state showing a crossover from boundary limited behavior at high temperature to a low temperature behavior that is different from that expected for bulk. Also shown are the calculated bulk behaviors: black long dashed line (22 bar), blue short dashed line (0 bar) for the bulk fluid thermal resistance in parallel with thermal boundary resistance. The black solid line (22 bar) and blue dot-dashed line (0 bar) show behavior expected for an isotropic distribution of point scatterers that give rise to a limiting mean free path of 1.1~$\mu$m. Open symbols show data just below $T_c$ (marked by vertical dashed lines in blue (0 bar) and black (22 bar)). The bulk calculations reference measured specific heat\cite{Greywall86SH}, thermal conductivity\cite{Greywall84TC} and thermal boundary resistance\cite{Cousins94}.}
\end{figure}

\section*{Results}
\subsection{Normal state measurements.}
Figures~\ref{fig::3_NormalState}(a,b) show the recovery time, $\tau$ and the extracted thermal resistance, $R_{th}$ (see Methods).
Measurements are shown at two pressures, 0 and 22 bar. At any temperature, the inelastic mean free path at low pressure is approximately three times longer than at the higher pressure. This should allow a study of the systematics of the crossover from bulk thermal conductivity to boundary limited scattering. However, at high temperatures (above 20 mK) a parallel conduction path into the IC chamber through the coin-silver walls dominates, complicating the crossover. Below $\sim$10~mK, the situation is simplified since transport through the channel dominates. Two sets of calculated curves are shown: long (22 bar) dashed and short (0 bar) dashed lines calculated by modeling the fluid as a bulk liquid, and a solid line (22 bar) and dash-dotted line (0 bar) representing added isotropic scatterers with a density sufficient to yield a mean free path of 1.1~$\mu$m.  In both cases a parallel conduction channel was added to model the conduction via cell walls. Vertical dashed lines mark the temperature of the superfluid transition at each pressure.

In Figure~\ref{fig::4_Kappa} we display the effective thermal conductivity, $\kappa_{\rm{EFF}}=lR_{th}^{-1}A^{-1}$ ( $l$, the channel length = 100 $\mu$m, $A$, the channel cross sectional area = 3 mm $\times$ 1.1 $\mu$m) below 10 mK, calculated from the measured thermal resistance and geometrical parameters of the channel. Below 10 mK, the conduction for the parallel thermal path (Kapitza resistance) is negligible and so the results represent the diffusive thermal conductivity of normal $^3$He in the channel. For both pressures, the thermal conductivity approaches a constant value. The behaviors expected for the bulk liquid ($\kappa_{\rm{EFF}} \propto T^{-1}$) and for isotropic scatterers ($\kappa_{\rm{EFF}} \propto T$) with a mean free path of 1.1~$\mu$m are shown as dashed and solid lines. Instead we find $\kappa_{\rm{EFF}}(T) \rightarrow$ constant as $T \rightarrow T_c$ in the normal state (where $\kappa (T_c)$ is determined to be 0.047$\pm$0.005 WK$^{-1}$m$^{-1}$ (0 bar), 0.024$\pm$0.003 WK$^{-1}$m$^{-1}$ (22 bar)).

\subsection{Superfluid state.} Below the superfluid transition temperature $T_c$, we carried out experiments at 0~bar, 0.62~bar and 22~bar. The lowest pressure (0 bar) was chosen because the inelastic mean free path in both the normal and superfluid states would be the longest. At the nearby pressure of 0.62 bar, $T_c$ is almost 10\% above its value at 0~bar and the inelastic mean free path at $T_c$ is already $\sim$20\% shorter than at 0 bar. At 22 bar, the inelastic mean free path will be much smaller than the other two pressures and there will be only a small temperature window in which the A phase will be present in the two bulk chambers near $T_c$. (Two superfluid phases are found in bulk $^3$He in zero magnetic field: the anisotropic A phase occupies the high pressure $P \ge$ 21.22 bar region near $T_c$, the isotropic B phase occupies the remainder of $P,T$ space\cite{Greywall85MCT}). Confinement of superfluid $^3$He in the channel of height $d = 1.1~ \mu m$ modifies this phase diagram\cite{Zhelev17NC,LevitinPRL19} and stabilizes the A phase over a significant temperature range at low pressures (to 0.7 $T_c$ (0 bar), 0.77 $T_c$ (0.62 bar), 0.82 $T_c$ (22 bar)).

The measured relaxation times, $\tau$, in the superfluid state are plotted in Figure~\ref{fig::5_SuperfluidTauKappa}(a) and the inferred thermal conductivity $\kappa_{\rm{EFF}} (T)$ in the channel is shown in Figure~\ref{fig::5_SuperfluidTauKappa}(b). The precision of the experiment did not reveal differences in $\tau$ through the expected A-B transitions in the channel. For comparison we show the results of model calculations (for bulk superfluid $^3$He-B) from Einzel\cite{Einzel84JLP}. We defer consideration of the differences in the responses at the three pressures to the discussion section.

\subsection{Anomalous response.} We now report on the anomalous thermal responses detected by the HEC fork. In this series of experiments we applied long duration pulses (between 100 to 300 s to deposit more heat) to the IC fork resulting in a change in temperature to the IC of order $\Delta T_{IC}$ $\sim$ 5-10\%. At 22~bar and well below $T_c$ we observed a small but immediate response in the HEC fork, decaying with a time constant similar to that for thermal relaxation in the IC. These pulses received at the HEC fork attain their maximum value at the end of the long duration pulse in the IC, indicating that the response of the HEC fork is proportional to the temperature excursion in the IC, $\Delta T_{IC}$.

The response of the HEC fork in the vicinity of $T_c$ at 0 bar is shown in Figure \ref{fig::6_NearTc}. The data demonstrates that the anomalous response requires the channel to be in the superfluid phase and cannot be due to crosstalk (which would give rise to visible signatures both above and below $T_c$). The magnitude of this response and the fact that is essentially immediate, and correlates with the temperature of the IC, indicates that the HEC fork is not registering a change in the equilibrium temperature of the HEC. Rather this is a ``local" response to the quasiparticle flux, initiated by heating of the IC. We note that the HEC fork is located near the channel mouth. See discussion for further details.

In Figure~\ref{fig::7_LocalTemp}(a-i) we show plots of the local temperature of the HEC fork ($T^{HEC*}/T_c$) at 0 bar (a, b, c), 0.62 bar (d, e, f), and 22 bar (g, h, i), compared to the temperature pulse in the IC for a selection of reduced temperatures. They show that the strongest anomalous response is seen at the lowest temperatures, and is significantly weaker at the highest pressure. At 22 bar no response in the HEC fork is seen for $T/T_c \ge$ 0.6, while at 0 bar the response persists to $T_c$ (Figures~\ref{fig::6_NearTc},
\ref{fig::7_LocalTemp}(a-c)
). Complete traces across the full temperature range in the superfluid of the observed change in $T^{HEC*},~T^{IC}$ after the application of pulses to the IC fork are shown for the 3 pressures measured in Supplemental Figure S 1.  We also show (in Supplemental Figure S 2) the evolution of the $Q$ in both HEC and IC forks following (comparable temperature excursion) pulses applied to the HEC fork. These results show that the anomalous heat flow is bi-directional.

\section*{Discussion}

At temperatures below 10 mK, heat flow from the IC directly into the coin-silver walls and thence to the nuclear stage can be neglected. Above this temperature, the measured thermal resistance can be modelled by parallel conduction through the channel and the boundary resistance and is consistent with a thermal boundary resistance scaling as $T^{-3}$ (see methods) as seen in Figure~\ref{fig::3_NormalState}. The inferred effective thermal conductivity $\kappa_{\rm{EFF}}$, corresponding to diffusive thermal transport in normal $^3$He in the 1.1 $\mu$m channel is seen to be nearly temperature independent below 10 mK at both 0 bar and 22 bar (Figure~\ref{fig::4_Kappa}).

At first sight this result appears anomalous and unexpected. The thermal conductivity, $\kappa$ of a Fermi liquid is proportional to the thermal mean free path $\lambda_{\kappa}$ (see introduction). In a bulk Fermi liquid in the absence of impurities, $\lambda_{\kappa} \propto T^{-2}$ yielding $\kappa \propto T^{-1}$ since $C_v \propto T$. Impurities lead to a temperature independent contribution to the mean free path\cite{Sauls00} and hence $\kappa \propto T$, and this might also be expected for boundary limited scattering (See also Methods). The onset temperature for such boundary limited scattering can be estimated from the bulk thermal mean free path $\lambda_{\kappa}T^2$ = 23.6~$\mu$m$\cdot$mK$^2$ at 0~bar and 7.36~$\mu$m$\cdot$mK$^2$ at 22~bar\cite{Greywall84TC}. At the lowest pressure we expect strong mean free path effects in the 1.1 $\mu$m channel below ~5 mK.

As shown in Figure~\ref{fig::3_NormalState}(b) and Figure~\ref{fig::4_Kappa}, the observed behavior for the thermal conductivity in our restricted flow channel at 0 bar (blue circles) clearly lies between these two extremes. At 22 bar (black triangles) the data lie well below the expected bulk behavior. However, the relatively small mean free path and the effect of the parallel Kapitza resistance result in the thermal conductivity at this pressure being similar to that for the impurity effect dominated behavior which has not transitioned to its $T$ dependence. The observed thermal conductivity at both pressures below 4 mK is temperature independent, suggesting an effective mean free path proportional to $T^{-1}$. While inconsistent with Fermi liquid theory, such an unusual temperature dependence of the mean free path has previously been inferred in studies of mass transport in $^3$He films over polished silver surfaces\cite{Casey04}, which found a momentum relaxation time ($\tau_{\eta}$) $\propto T^{-1}$ at temperatures below 100 mK. This result was interpreted\cite{Sharma11PRL} in terms of quasiclassical interference between bulk and boundary scattering channels, as earlier proposed for thin metal films with rough surfaces\cite{MeyerovichPRB1998,MeyerovichPRB2002,MeyerovichPRB2010}. This theory has subsequently been extended to thermal transport\cite{Sharma2018} yielding a constant value of $\kappa$ over a wide temperature range for surface roughness of 3 nm rms with fractal correlations. This is likely to be beyond the upper limit for roughness of our glass substrate\cite{Zhelev18RSI,Sharma2018}. A more likely source of scattering is the presence of trapped charges at the surface of the silicon. Such charges (due to dangling bonds) would induce local density variations that would create random scattering potentials, mimicking surface roughness.


Our results in the superfluid state (Figure~\ref{fig::5_SuperfluidTauKappa}(
b)) constitute a measurement of the diffusive thermal conductivity in the absence of hydrodynamic heat flow. A strong contribution from thermal counterflow\cite{Kleinberg74,Johnson75} has been observed in bulk superfluid $^3$He. Second sound\cite{Kojima} has also been observed in bulk superfluid $^3$He. Under confinement the normal fluid is expected to be clamped and so the thermal counterflow contribution should be eliminated. We carefully estimate the hydrodynamic thermal conduction and confirm that it is negligible in our slab in the presence of strong slip effects. We find (in Supplementary Note S-1) that comparable diffusive and hydrodynamic contributions to thermal conduction would arise if $d$, the confinement height were of order 100 $\mu$m to 1 mm, due to the $d^{-3}$ contribution to the impedance $Z$ and taking into account the confinement dependence of the effective viscosity.

There has been no prior systematic experimental study of the diffusive thermal conductivity of superfluid $^3$He. However, the theory of spin independent transport has been developed for bulk\cite{Einzel1978JLTP,Wolfle1978JLTP,Einzel84JLP} and is parameterized by a series of well-defined relaxation times and a pressure dependent scattering parameter, which for the thermal conductivity is referred to as $\lambda_1^-$. Over a reasonable range of possible values for $\lambda_1^-$ ($ 0.9 \leq \lambda_1^- \leq 2.0$) the thermal conductivity at low temperatures, ($T/T_c$ = 0.4), relative to that at $T_c$ varies by at most a factor of two\cite{Einzel84JLP}.

As previously discussed we find that at both low pressure and 22 bar, under strong confinement, $\kappa_{\rm{EFF}}(T_c)$ is reduced below the bulk value at variance with standard theory of the normal state. In Figure \ref{fig::5_SuperfluidTauKappa}(b) we compare our measurements at 22 bar with the theory for bulk diffusive thermal conductivity. The thermal conductivity at 22 bar shows a weak maximum and then decreases as the temperature is lowered below $T_c$. At low pressure the thermal conductivity increases before plateauing at approximately twice the value at $T_c$.  In the light of these results, a full calculation of the diffusive thermal conductivity under different degrees of confinement is highly desirable. For channels with intermediate confinement of order the mean free path, this would extend the results from mass transport\cite{EinzelParpia87} to thermal transport, and include a treatment of surface slip and surface Andreev scattering into the extreme Knudsen regime. Under conditions of strong confinement for which the channel height is comparable to the superfluid coherence length (typically below 1 $\mu$m at zero bar), a full quasi-classical calculation incorporating the contributions of surface bound states is required.

\begin{figure}[ht!]
\begin{center}
\includegraphics[width=0.8\textwidth]{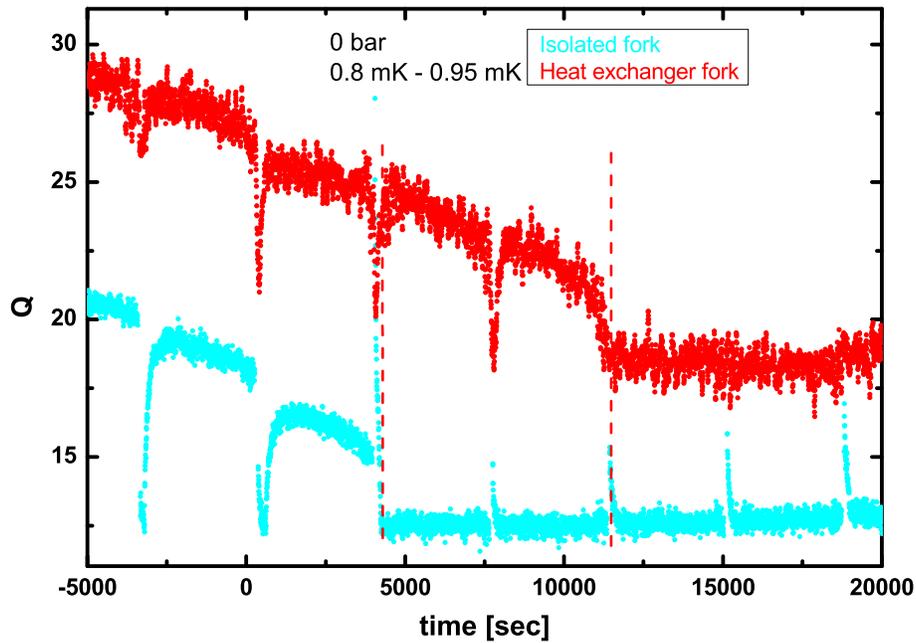} 
\end{center}
\caption{\label{fig::6_NearTc}
The $Q$ as a function of time for the two forks (cyan IC, red HEC) near $T_c$ taken at 0~bar. Pulses applied were separated by 60 minutes and the nuclear stage was warmed up at a constant rate around 10~$\mu$K$\cdot$hr$^{-1}$. When both the IC, HEC are in the superfluid state (first three pulses) a strong response is seen in $Q_{HEC}$. As the IC passes through its $T_c$, a pulse applied evokes a response in $Q_{HEC}$. Even when the IC is in the normal state (the response in $Q$ is reversed) a strong response is seen in $Q_{HEC}$. Only when both the HEC and IC are in the normal state is there no anomalous response seen in $Q_{HEC}$ to a heat pulse in the IC. We hypothesize that when the IC is in the normal state and the HEC is in the superfluid state, the channel is more tightly linked to the HEC and in the superfluid state.}
\end{figure}

\begin{figure}[ht!]
\begin{center}
\includegraphics[width=\textwidth]{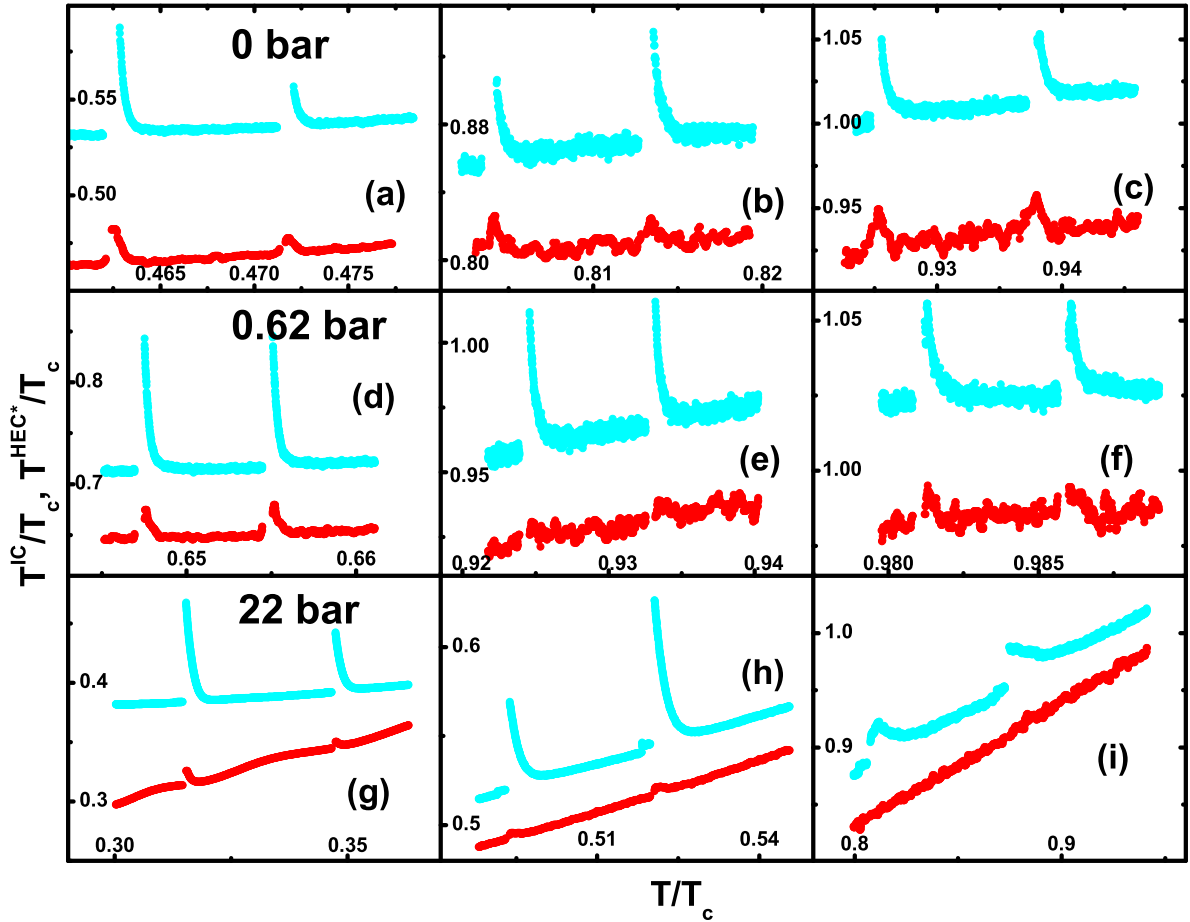} 
\end{center}
\caption{\label{fig::7_LocalTemp}
  The 3 $\times$ 3 panels (a-i) show the local temperature $T^{HEC*}/T_c$ (red dots) compared to $T^{IC}/T_c$ (cyan dots) [offset upward by 0.05 $T^{IC}/T_c$] for clarity against $T/T_c$. 
  The top row shows three representative sets of pulses at 0 bar and near 0.5 $T/T_c$ (a), 0.8 $T/T_c$ (b) and 0.94 $T/T_c$ (c). The middle row (d, e, f) shows results for 0.62 bar, and the lower row shows results at 22 bar (g, h, i). Two different pulse durations were used accounting for differences in the initial temperature rise observed. Data obtained while cooling exhibits the same behavior.}

\end{figure}

We now turn to the anomalous thermal response in the HEC fork  (Figures~\ref{fig::6_NearTc}, \ref{fig::7_LocalTemp}) seen at low pressure and below 0.6 $T_c$ at 22 bar. The association of this conduction with superfluidity in the channel is best seen in Figure~\ref{fig::6_NearTc} where the signal in the HEC fork vanishes in the normal state. The response is attributable to a quasiparticle flux incident on the HEC fork driven by the fountain pressure generated in the IC (accompanying the elevated temperature in the IC) following the heating pulse. In Supplemental Note S-1 we calculate the hydrodynamic heat flux, dQ/dt and find that it cannot account for the observed temperature increase in $T^{HEC*}$. The hydrodynamic heat flux represents only a small fraction of the heat conducted by diffusive processes and would be too small to measure. Further, if the temperature rise was indicative of the temperature of the HEC, it would require a greater amount of heat input than that deposited into the IC. Such a heat flow would have a long rise time governed by the heat capacity of the HEC. Thus the registered response in the HEC fork ($T^{HEC*}$) is ``local", and does not reflect an increase in the equilibrium temperature of the HEC.

\begin{figure}[ht!]
\includegraphics[width=\textwidth]{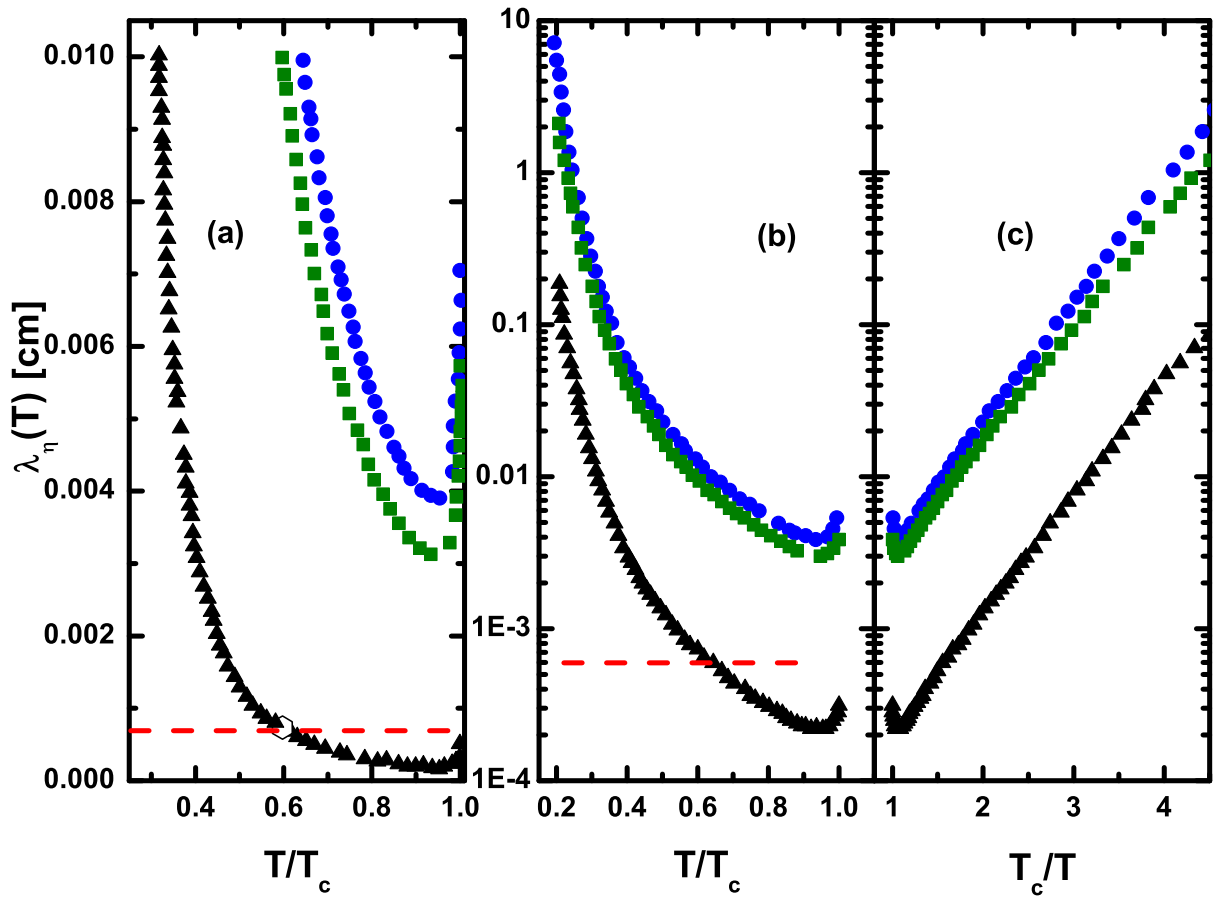} 
\caption{\label{fig::8_MFP}
(a) The viscous mean free path, $\lambda_{\eta}$ below $T_c$ as a function of $T/T_c$ and $T_c/T$ for 0 (blue circles), 0.62 (green squares) and 22 (black triangles) bar pressures. Note the strong increase in $\lambda_{\eta}$ from a minimum just below $T_c$. (b) and (c) show the viscous mean free path in cm on a logarithmic scale. At 0.6 $T/T_c$ for 22 bar, the mean free path is approximately 6~$\mu$m (dashed red line). }
\end{figure}

We observe this anomalous transport only when the mean free paths are long. In Figure~\ref{fig::8_MFP} we show a plot of the viscous mean free path as a function of reduced temperature $T/T_c$\cite{EinzelParpia87}. For 0 bar at $T_c$ the viscous mean free path, $\lambda_{\eta}$, is $\sim$ 72 $\mu$m. It then decreases rapidly by about a factor of 2 below $T_c$ before rising exponentially at low temperatures, ensuring that the channel is well within the Knudsen regime at all temperatures below $T_c$. We believe that the long mean free path allows the normal fluid to slip and flow in response to the rise in fountain pressure at all temperatures at 0 bar. At the intermediate pressure (0.62~bar) the received signal in the HEC fork is present at $T_c$, then following trends in the mean free path, it gets weaker below $T_c$ before growing as the temperature is further lowered. This temperature dependence can be seen in Figure~\ref{fig::7_LocalTemp}(d,e,f) and in the continuous trace data (and inset) shown in Supplemental Figure S 1. The smallest response at 0.62 bar is aligned with the location of the minimum in the viscous mean free path (Figure \ref{fig::8_MFP}). At 22 bar the anomalous response is only observed at $T/T_c \leq $0.6 and supports the hypothesis that the anomalous heat conduction mechanism appears only when the mean free path is sufficiently large ($\lambda_{\eta} \ge$ 6 $\mu$m, Figure \ref{fig::8_MFP}).

The hydrodynamics in this long mean free path and restricted geometry regime has not been explored theoretically or experimentally in the context of superfluid $^3$He. Following a heat pulse in the IC, we associate the observed excess damping in the HEC fork with an increased local temperature, $T^{HEC*}$ due to an incident quasiparticle flux. At these large Knudsen numbers, in an analysis of hydrodynamic flow, the normal fluid velocity is constant across the height of the channel (plug flow). The flow velocity will thus be affected by surface quality and Andreev scattering processes at the surface. Nevertheless, it is unlikely that quasiparticles would transit through the intervening bulk superfluid to interact with the HEC fork. In our highly confined channel ($d/\xi_0 \sim 15$ at P = 0 bar, $d$ being channel height, $\xi_0$, the coherence length) we should consider the influence of surface bound excitation states\cite{ZhangPRB1987,ThunebergPhysica1992,Nagato1998,VorontsovPRB2003}. The dynamics of these excitations and the interplay with bulk quasiparticle excitations under non-equilibrium conditions is not fully understood, and there have been recent studies in which they may play an important role\cite{LeePRL2017,ZmeevNatPhys2016}. We speculate that the surface excitations respond to the fountain pressure between the two chambers and may flow with little dissipation over distances much longer than the inelastic mean free path for bulk quasiparticles. If at the throat of the channel, conditions are satisfied so that these surface excitations can be injected into the bulk, they may be responsible for the local transient response, $\Delta T^{HEC*}$ seen by the HEC fork. There are indications that the flow of excitations may be subject to limitation above some threshold flux (See Supplemental Note S-2, Supplemental Figure S 4).

In conclusion, we have made a study of thermal transport through a 1.1~$\mu$m tall cavity in both the normal and superfluid phases of $^3$He. There are three principal findings.

First, the effective thermal conductivity of normal $^3$He under this strong confinement is temperature independent below 10 mK. Consequently the magnitude of the conductivity at $T_c$ is quantitatively different from that in bulk. The temperature independence can be understood in terms of an effective thermal mean free path that varies as $T^{-1}$, rather than $T^{-2}$ (bulk inelastic mean free path) or constant (boundary limited scattering). This is qualitatively consistent with previous studies of mass transport in thin films, that is accounted for by a theory of interference between inelastic scattering within the film and elastic scattering arising from an effective disorder potential originating from surface roughness.

Second, the relatively weak temperature dependence of the diffusive thermal conductivity in the superfluid state, relative to its value at $T_c$, is similar to that calculated for bulk liquid. This result motivates further measurements of thermal transport in a slab-like cavities, sufficiently confined to make hydrodynamic heat flow small compared with diffusive heat flow, but as large as possible to minimise the effects of surface slip, and minimise the contribution of surface states. By contrast, the height of the present cavity was chosen to approach the superfluid coherence length at the lowest pressures. In this case a full quasi-classical calculation incorporating the contribution of surface excitations to the diffusive thermal transport is highly desirable.

Third, despite the fact that according to our estimates, the overall thermal transport should be dominated by diffusive thermal transport, we observe a thermal response driven by the fountain pressure difference between the two chambers, in either direction. As we discuss, this may be evidence of quasi-ballistic thermal transport due to the surface excitations.

With attainable improvements in the precision of thermometry, this work opens the prospect of a variety of thermal transport studies of topological superfluid $^3$He under strong confinement at length scales comparable to the superfluid coherence length. Thermal transport should be sensitive to the presence of interfaces in the superfluid, either those arising spontaneously as in the spatially modulated superfluid, or those engineered by steps in cavity height, since confinement controls the stable superfluid order parameter. The detection of surface, edge and interface excitations by thermal transport, and the thermal Hall effect and edge currents in topological superfluid $^3$He should act as a benchmark for similar studies of putative topological superconductors.

\section*{Methods}

\subsection{Thermal Conduction Channel Construction.}
We fabricated the entire assembly (Isolated Chamber, thermal conduction channel holder, Heat Exchanger Chamber) out of coin silver (Figure~\ref{fig::1_Details of Experimental Cell}, Supplemental Figure S 5). This minimized time dependent heat leaks into the $^3$He because the walls were thermally well anchored to the nuclear stage.   The 5~mm $\times$ 5~mm  silicon chip (Figure~\ref{fig::1_Details of Experimental Cell}(a)) that comprises the nanofabricated channel and establishes the thermal impedance between the two chambers for $^3$He was made at Cornell$'$s Nanofabrication facility. The process flow is similar to that detailed elsewhere\cite{Zhelev18RSI}. After patterning of the silicon, a matching square piece of highly polished sodium doped glass (Hoya SD-2) was bonded to the silicon. After bonding, the edges of the silicon and glass that were parallel to the heat flow were rounded off using a high speed Dremel tool and carborundum bit. The cavity was mounted (using Trabond epoxy) into a coin silver holder that had walls that were machined to a 0.15~mm thickness (Supplemental Figure S 5(b)). The rounded corners distribute stress on the silicon and glass components during thermal cycling. The thin coin-silver of the holder also enabled movement of the metal with the epoxy and silicon-glass cavity so as to accommodate thermal contraction on cooling. A dummy cavity (without a through pathway) was cycled repeatedly to liquid nitrogen temperatures and proved to be leak tight post-cycling. The design is such that there should be no differential pressure across the large faces of the cavity, thus no additional pressure dependent bowing should be present to alter the cavity dimensions as the pressure is varied\cite{Zhelev17NC}.

\subsection{Fork operation and pulses.}

The forks were driven using a constant voltage signal at a level small enough so that no drive dependent heating was observed. The detection was via a voltage preamplifier connected to one tine of the fork while the drive was applied to the adjacent tine. We estimate from the energy deposited during heat pulses and the ratio of drive voltages that the ambient heating due to operation of the forks is $\sim$ 0.1~pW near 1~mK.  The preamplifiers for each fork had their 6~dB/octave filters set at 10~kHz and 100~kHz.  In order for the feedback loop operate well at low temperatures (where the $Q$ is low, approaching 10 at $T_c$ at 0~bar for a resonant frequency $\sim$34~kHz) we measured (at 20 mK) and fitted the background (non-resonant signal) over a wide frequency range (10~kHz to 60~kHz) using a 5th order polynomial after excluding the region of the resonance. After subtraction of the non-resonant background signal, the inferred $Q$ was reliably indicative of the temperature of the liquid. Further sweeps were also carried out at lower temperatures below 3~mK where the $Q$ was lower to obtain better fits and identify and compensate for a small temperature dependent background. After the background was well fit, we could carry out a calibrating sweep at intermediate temperature (typically 10~mK) where the $Q$ was about 200, to establish the conversion from peak amplitude to $Q$ and then measure the temperature dependent real and imaginary components of the response while driving the forks at a constant frequency. If the entrained mass caused a shift in the resonant frequency that exceeded 10\% of the linewidth, we recomputed the resonant frequency and altered the drive frequency to coincide with the center frequency. Smooth responses of the real and imaginary components of the recorded signal across these re-balances assured us that the fits were accurate. The technique allowed us to track the $Q$ of the forks across $T_c$ at 0 bar, where the viscosity is largest.

Pulses were applied by increasing the drive voltage above ambient by a factor of 10. We could apply more heat as needed by increasing the duration of the high drive. During the high drive state, it was not possible to track the resonant frequency of the fork or its $Q$. Therefore we turned off the frequency re-balance component of the program and operated at a fixed drive frequency during the pulse. However, we forced a re-balance of the forks prior to applying the pulse, so that they would both be operating close to their individual resonant frequency during the recovery after the pulse. The re-balance is visible as a small discontinuity prior to the application of the pulse in the inferred $Q$ vs time shown in Figure~\ref{fig::2_Heat Pulses}. In practice we found the $Q$ to be more robust against any background corrections so the temperature was monitored using the $Q$.

\subsection{Calculation of Thermal Resistance.}
The measured $\tau$ values in the normal state were converted to an effective thermal resistance by evaluating the heat capacity of the isolated chamber using the interpolations provided in reference.\cite{Greywall86SH} We used the relationship $\tau$=$R_{th}C$, where $R_{th}$ and $C$ are the effective thermal resistance to the IC and heat capacity of the $^3$He in the IC respectively.
The thermal conductivity of $^3$He is relatively poor at high temperatures because the excitation density in this regime leads to a short mean free path\cite{DyPethick686}. In our arrangement, a parallel path for heat transport becomes significant above $\sim$ 10~mK, through the Kapitza boundary resistance, $R_K$\cite{Cousins94}. The boundary resistance varies as $T^{-3}$, resulting in a crossover around 10-20 mK from the surface dominated to the channel dominated resistance. The actual behaviour (see Figure~\ref{fig::3_NormalState}(b)), especially at 22 bar, does not follow the $T^{-3}$ power law, for two reasons. First, the measured Kapitza resistance for a sheet obeys a power law that is closer to $T^{-2.5}$ for this temperature range\cite{Cousins94}. Second, a portion of the surface area in the IC is in the form of two closely fitting cylinders. The effective area of the cylinders is modified by the conductivity of the $^3$He that fills the gap between the cylinders: the area participating in heat flow decreases as the temperature increases due to the temperature variation of the conductivity of the $^3$He.

\begin{figure}[ht!]
\begin{center}
\includegraphics[width=0.8\textwidth]{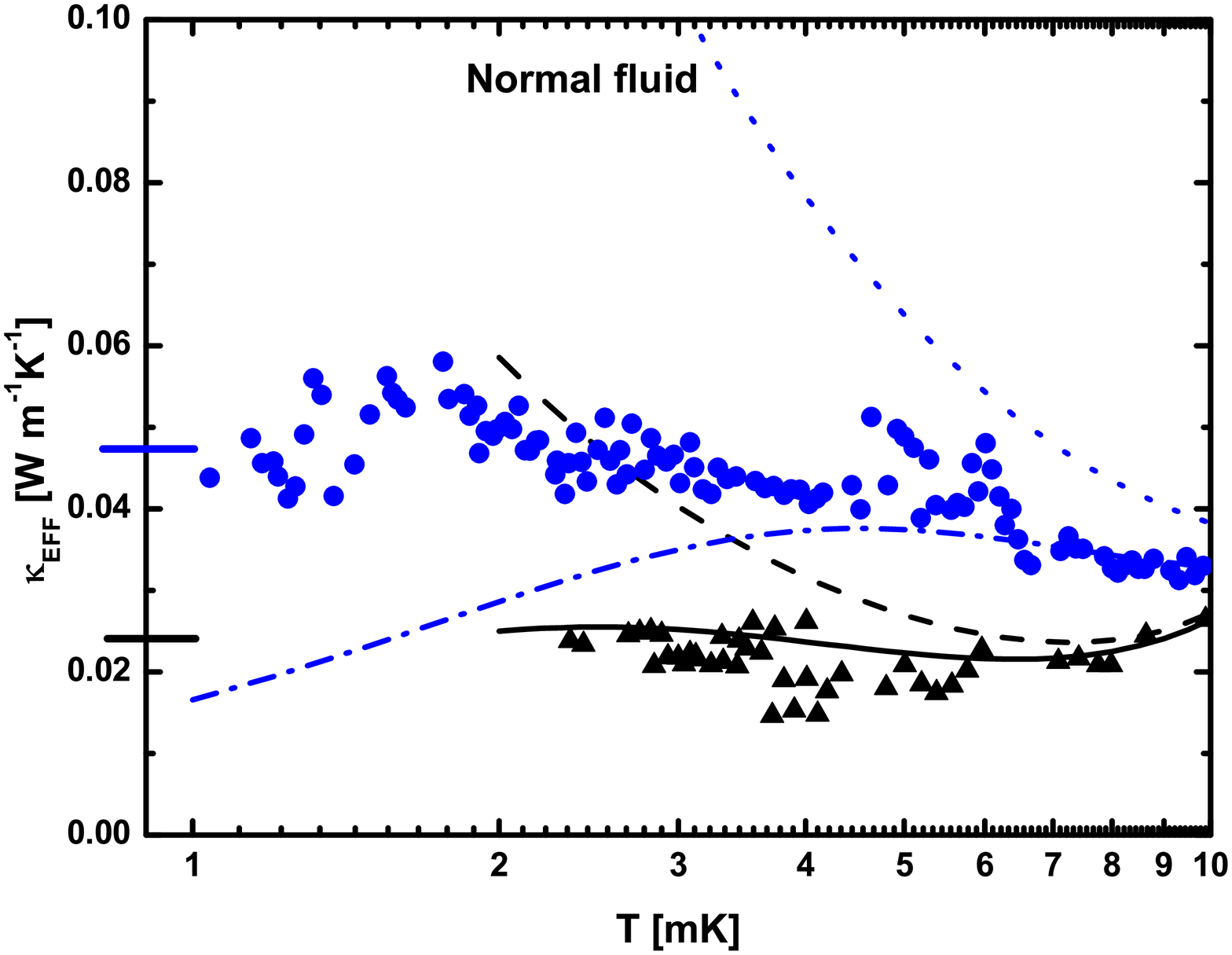} 
\end{center}
\caption{\label{fig::4_Kappa}
  The thermal conductivity $\kappa_{\rm{EFF}}(T)$ at 0~bar (blue circles) and 22~bar (black triangles) below 10 mK in the normal state calculated from geometrical parameters and $R_{th}(T)$. Also shown are the calculated bulk behaviors (short dashed blue line (0 bar), long dashed black line (22 bar) exhibiting the $T^{-1}$ dependence of inelastic scattering of Bogoliubov quasiparticles. The dot-dashed blue line (0 bar) and solid black line show the thermal conductivity expected for a distribution of point scatterers that give rise to a mean free path of 1.1~$\mu$m. The horizontal lines define the values for $\kappa_{\rm{EFF}}(T)$ at 0 bar (blue 0.047$\pm$0.005 WK$^{-1}$m$^{-1}$), and 22 bar (black 0.024$\pm$0.003 WK$^{-1}$m$^{-1}$) as $T_c$ is approached in the normal state.
  }
\end{figure}

\begin{figure}[ht!]
\begin{center}
\includegraphics[width=0.6\textwidth]{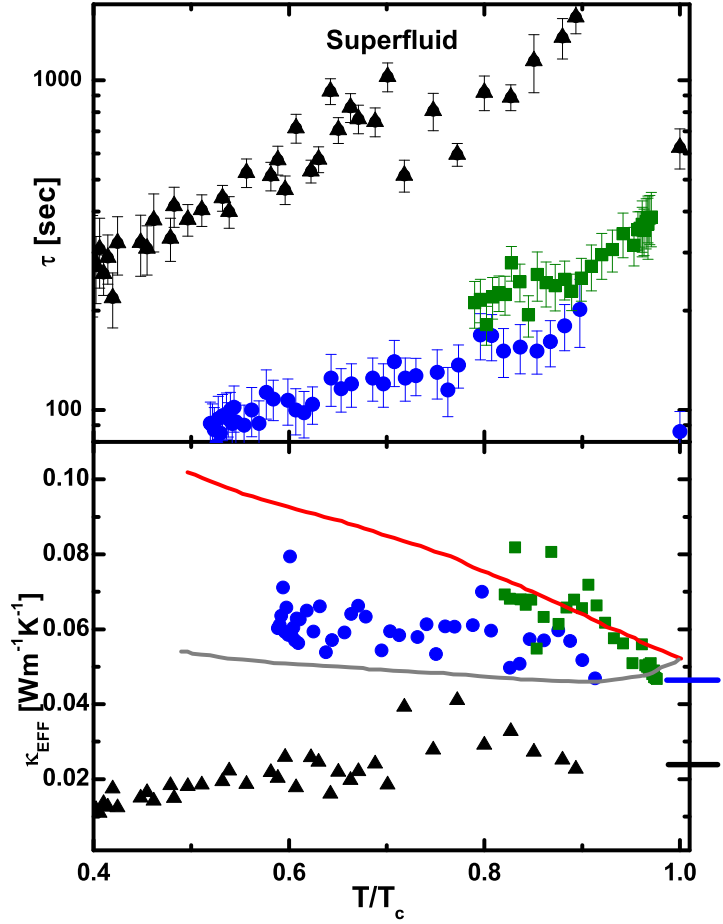} 
\end{center}
\caption{\label{fig::5_SuperfluidTauKappa}
  The measured thermal relaxation times (a) and calculated effective thermal conductivities, $\kappa_{\rm{EFF}}$ (b) at 0~bar (blue circles), 0.62~bar (green squares) and 22~bar (black triangles) below the superfluid transition temperature $T_c$. The bulk calculations reference calculated thermal conductivity\cite{Einzel84JLP} with $\lambda^-_1$ = 2 (grey solid line), = 0.9 (red solid line). The  (0 bar) and black (22 bar) horizontal lines mark the limiting ($T\rightarrow T_c$) value of $\kappa_{\rm{EFF}}$ in the normal state (see Figure \ref{fig::4_Kappa}). In the bulk and under confinement, $\kappa_{\rm{EFF}}$ is only weakly temperature dependent. No calculations of $\kappa (T/T_c)$ exist in the literature for 0 bar.} 
\end{figure}

In Figures~\ref{fig::3_NormalState}(a,b), \ref{fig::4_Kappa}, we also include the expected behavior (solid and dot-dashed lines) for the thermal conductivity of samples with similar geometry to that studied whose resistance is characteristic of a uniform distribution of elastic scatterers spaced to yield a 1.1~$\mu$m elastic scattering length.  We modify $\kappa$ = $1/3 (C_v/V)v_F^2\tau_{\kappa}$ by replacing $\tau_{\kappa}$ with an effective scattering time, $\tau_{\rm{eff}}$ given by Mathiessen's rule ($\tau^{-1}_{\rm{eff}} = \tau^{-1}_{\rm{el}}  +\tau^{-1}_{\rm{in}}$). Thus $\tau_{\rm{eff}}$ approaches a constant when the quasiparticle scattering time, $\tau_{\rm{in}}$, exceeds the impurity scattering time, $\tau_{\rm{el}} = 1.1~\mu m/v_F$. The resulting impurity dominated thermal conductivity thus varies as $T$.

\begin{addendum}
 \item We acknowledge input from Prof. J.A.~Sauls including access to preliminary calculations and thank Prof. A. Golov for providing a copy of the thesis of Wellard. We also acknowledge a helpful exchange with V. Ngampruetikorn regarding contributions of surfaces to thermal conduction. This work was supported at Cornell by the NSF under DMR-1708341 (Parpia), PHY-1806357 (Mueller), in London by the EPSRC under EP/J022004/1. In addition, the research leading to these results has received funding from the European Union’s Horizon 2020 Research and Innovation Programme, under Grant Agreement no 824109. Fabrication was carried out at the Cornell Nanoscale Science and Technology Facility (CNF) with assistance and advice from technical staff. The CNF is a member of the National Nanotechnology Coordinated Infrastructure (NNCI), which is supported by the National Science Foundation (Grant NNCI-1542081).

 \item[Correspondence] Correspondence and requests for materials should be addressed to J.M.P. \newline (email:  jmp9@cornell.edu).

 \item[Author contributions] Experimental work, analysis and presentation was principally carried out by D.L. with early contributions by A.E. assisted by M.T and J.W. with further support from E.N.S. and J.M.P.. N.Z had established most of the routines for the phase locked loop operation of the quartz fork for earlier experiments. E.M provided general guidance on thermal conductivity issues and significantly contributed to the data analysis protocols and the writing of the manuscript, and N.Z and T.S.A. established and carried out the nano-fabrication protocols. D.E. calculated the viscous and thermal mean free paths. J.M.P. supervised the work and J.M.P. and J.S. had leading roles in formulating the research and writing this paper. All authors contributed to revisions to the paper.

 \item[Data Availability] The data that supports this study will be made available through Cornell University e-commons data repository at https://doi.org/10.7298/4fhq-e356. 

\item[Competing Interests]
The authors declare that they have no competing financial interests.
\end{addendum}

\pagebreak
\pagebreak
\pagebreak
\pagebreak
\pagebreak

\section*{References:}
\bibliographystyle{naturemag}
\bibliography{sample}

\end{document}